# Open Access Dataset for Electromyography-based Multi-code Biometric Authentication

Ashirbad Pradhan, Jiayuan He, Member, *IEEE*, and Ning Jiang*, Senior *Member, IEEE*

***Abstract*— Recently, electromyogram (EMG) has been proposed as a novel biometrics for addressing some key limitations of current biometrics. EMG signals are inherently different for individuals (biometrics), and they can realize multi-length codes or passwords (for example, by performing different gestures). However, current EMG-based biometric research has two critical limitations: 1) a small subject pool, compared to other more established biometric traits, and 2) limited to single-session data sets. In this study, forearm and wrist EMG data were collected from 43 participants over three different days with long separation (Day 1, 8, and 29) while they performed static hand/wrist gestures. The multi-day biometric authentication resulted in a median EER of 0.017 and 0.025 for the forearm and wrist setup, respectively. These results demonstrated the potential of EMG-based biometrics in practical applications. The large-sample multi-day data set would facilitate further research on EMG-based biometrics and other EMG gesture recognition-based applications.**

## I. INTRODUCTION

BIOMETRICS has become an integral part of current authentication systems, and has found application in consumer electronics, public security, and private security. These biological and behavioral traits have been utilized to identify an individual or verify an individual's identity. Conventional biometrics such as fingerprints and facial scans have been widely used in smartphones and laptops in our daily lives. However, with technological advancements, there are increasing risks of leakage of biometric data as well as artificial regeneration (also termed spoof), which can lead to the identity theft of an individual. Recently novel biometric traits based on bio-signals, such as the electrocardiogram (ECG), electroencephalogram (EEG), and electromyogram (EMG) have been shown to be more resilient to spoofing than the conventional biometric traits [1]. Among these, surface EMG has been traditionally used in gesture recognition-based research, specifically for prosthetic control, where extensive investigation demonstrated EMG-based gesture control has poor cross-user transference performance [2]. In fact, a calibration-free EMG-based gesture recognition system that does not need new user training has been an elusive goal in myoelectric control literature, which suggests that there exist inherent individual differences in surface EMG signals. And this is precisely a biometrics trait. Indeed, multiple recent studies have substantiated EMG as an accurate biometric trait [3-6]. In this context, EMG has an inherent dual-property: gesture recognition and biometrics, providing it with a distinct and important advantage over other biometric traits. On the one hand, it is more covert than the traditional traits, such as fingerprint, and less likely to be compromised and spoofed. On the other hand, it enables the user to set customized gestures as passcodes for enhanced security, just like a user-defined password, which is not possible with EEG and ECG. Our recent study on multi-code EMG biometrics has provided a framework for the fusion/combination of these codes and to facilitate such a dual-mode (password and biometrics) authentication system [7]. Another recent study used a multi-code framework to incorporate password-based security and achieved similar results for biometric authentication [8].

### A. *State-of-the-art in EMG-based Biometric Authentication*

There are generally two common biometric modes: Verification and Identification [9]. In the Verification mode, the biometric system grants or rejects the access request of the presenting user (claimant) by comparing the presented biometric data to the template stored in the database. In this case, the presumed identity of claimant is *a priori*. In the Identification mode, however, the presumed identity of the claimant is unknown and the biometric system has to identify the most likely identity from the database based on presented biometric information. In the current study, we focus on the verification mode, or commonly termed authentication mode, as it is widely used in daily life. Some of the studies have reported a high biometric authentication performance (>95%) [5, 6, 8, 10, 11]. The number of hand gestures in these studies ranged from 1 to 34. The number of EMG channels varied from as low as one channel to as high as 256 channels. While most of the studies had less than 25 subjects, only three studies had a larger number (>40) of subjects [5, 12, 13], more appropriate in the biometric context. In addition, it is important to investigate the multi-session and multi-day robustness of biometric traits. However, these studies were limited to data acquired within only one session or one day. A few studies with a small subject pool (<22), with data from a two-day protocol [3, 8, 14, 15]. Only one study had five subjects with a four-day data collection protocol [16]. It has been established in the EMG processing literature, that in a multi-session protocol spreading across days, non-stationary factors including electrode shifts, sweat,

This work was supported by the Natural Sciences and Engineering Research Council of Canada (Discovery Grant: 072169). AP, JYH and NJ are with the Engineering Bionics Lab, Department of Systems Design Engineering, Faculty of Engineering, University of Waterloo, Waterloo, Canada (* Corresponding author: ning.jiang@uwaterloo.ca)

and dry skin, and physical conditions will affect the accuracy and consistency of EMG processing system [17]. Therefore, the multi-day performance with a sufficiently large subject pool is a crucial step for validating the effectiveness of EMG as a biometric trait.

TABLE I
MULTI DAY EMG DATABASES

| Database | Subjects | Days | Gestures | Channels |
|---|---|---|---|---|
| Megane Pro [23] | 10 | 5 | 7 | 14 |
| CapgMyo [18] | 10 | 2 | 8 | 128 (HD) |
| CSL-HDEMG [19] | 5 | 5 | 27 | 192 (HD) |
| ISRMyo-I [24] | 6 | 10 | 10 | 16 |
| DS-DataSet [21] | 42 | 2 | 7 | 6 |
| putEMG [20] | 44 | 2 | 8 | 24 |
| Hyser [22] | 20 | 2 | 34 | 256 (HD) |
| GrabMyo (current) | 41 | 3 | 16 | 28 |

### B. Open-access EMG Datasets

There are several open-access multi-session EMG databases in the literature, which are summarized in Table I. Some open-access databases of multi-day EMG recordings of forearm muscles are available which could facilitate the multi-day performance investigation [18-22]. Two databases with large subject pool (>40) involved two days of data collection [20, 21]. Some studies used as low as 6 electrode channels [21], while some utilized a high-density (HD) EMG setup [18, 19, 22]. Only three studies involved a higher number (>2) of data collection days, but the number of subjects was less (<11) [18, 23, 24]. To further explore the suitability of EMG biometrics as an accurate and robust industrial application, it is imperative to have a database with larger subject pool sizes, recorded across multiple days.

In the current study, we provide an open-access Gesture Recognition and Biometrics ElectroMyography (GrabMyo) Dataset. The presented dataset is the largest EMG dataset in terms of the total number of recording sessions (43 subjects x 3 days = 129 recording sessions). A 2048 Hz biosignal amplifier was used. In addition to EMG-based biometrics research, the dataset also provides a valuable resource for EMG-based gesture recognition research, particularly for improving algorithms' robustness in a multi-day scenario. Further, the current database has two electrode positions: forearm and wrist. This configuration will facilitate the research and development of industry-grade wearable wrist bands and bracelets for biometric authentication and gesture recognition applications [13].

For the scope of this paper, the standard biometric analysis was performed on the GrabMyo dataset, and the benchmark results of biometric authentication were reported. We have made our dataset publicly available on Physionet [link available upon request] and IEEE Dataport [link available upon reques].

TABLE II
DATABASE SUMMARY

| Participant Characteristics | Values |
|---|---|
| # Males | 23 |
| # Females | 20 |
| # Right-handed | 38 |
| # Left-handed | 5 |
| Age (years) | 26.35 ± 2.89 |
| Forearm length (cm) | 25.15 ± 1.74 |
| Forearm circumference (cm) | 24.10 ± 2.27 |
| Wrist circumference (cm) | 16.18 ± 1.21 |

The forearm length is measured from the olecranon process to the ulnar styloid. The forearm circumference is measured at a distance one-third from the elbow joint. The wrist circumference is measured 1 cm away from the ulnar styloid process

## II. METHODS

### A. Participants

We recruited 43 healthy participants (23 M, 20 F) for the study. The average age was 26.35 ± 2.89, and the average forearm length (measured from the styloid process on the wrist to the olecranon on the elbow) was 25.15 ± 1.74 cm. More details about the dataset and the participant characteristics are reported in Table II. Before the experiment, the participants were informed of the procedures and signed an informed consent form. The experiments were conducted following the Declaration of Helsinki and the research protocol was approved by the Office of Research Ethics of the University of Waterloo (ORE# 31346).

### B. Acquisition Setup

The experimental setup consisted of a PC and a monitor mounted on a desk, 0.75 m in front of a height-adjustable chair. The EMGUSB2+ (OT Bioelettronica, Italy), a commercial amplifier, was used for acquiring the sEMG signals. The signals were bandpass filtered between 10 Hz and 500 Hz, with a gain of 500, and then sampled at of 2048 Hz.

Prior to the experiment, the participant's forearm length is measured as the distance between the olecranon process and the ulnar styloid process. The forearm circumference is measured at one-third of the forearm length from the olecranon process. The wrist circumference is measured at 2 cm away from the ulnar styloid process. After taking these measurements, the electrodes are placed on the forearm and wrist. For the forearm electrode placement, sixteen monopolar sEMG electrodes (AM-N00S/E, Ambu, Denmark) were placed in the form of two rings, each consisting of eight electrodes equally spaced around the forearm, forming eight bipolar pairs. The center-to-center distance between the two rings was maintained at 2 cm. For the wrist electrode setup, twelve monopolar sEMG electrodes of the same type as the forearm rings were placed in the form of two rings, each consisting of six electrodes equally spaced around the wrist and forming six bipolar pairs. The center-to-center distance between the two rings was maintained at 2cm, similar to the forearm setup. Therefore, a total of 28 monopolar sEMG electrodes were used for each session. A detailed pictorial representation is provided in Fig. 1. To maintain consistency of the positions of the electrodes across all participants, the first electrode in each ring (total rings = 4) was anatomically positioned on the centerline of the elbow crease as shown in Fig. 1 [6, 11].

TABLE III
GESTURE LIST

| Gesture | Description | Gesture | Description |
|---|---|---|---|
| 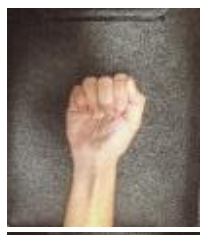 | Lateral prehension (LP) | 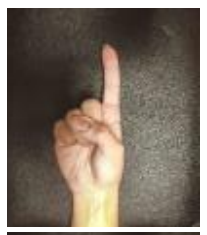 | Index finger extension (IFE) |
| 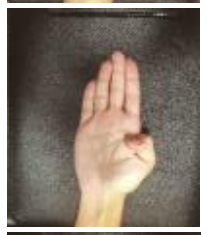 | Thumb adduction (TA) | 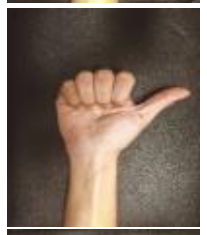 | Thumb extension (TE) |
| 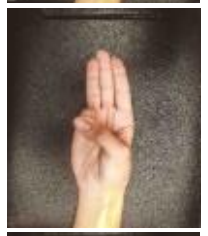 | Thumb and little finger opposition (TLFO) | 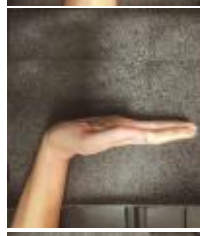 | Wrist flexion (WF) |
| 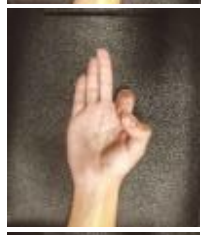 | Thumb and index finger opposition (TIFO) | 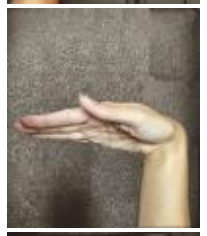 | Wrist extension (WE), |
| 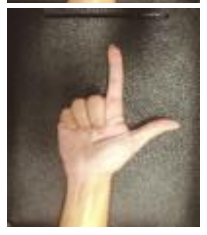 | Thumb and little finger extension (TLFE) | 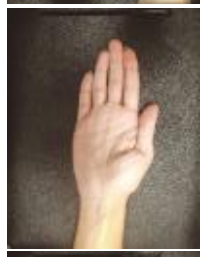 | Forearm pronation (FP) |
| 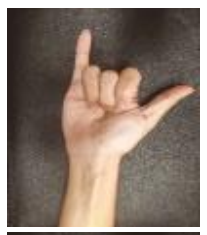 | Thumb and index finger extension (TIFE) | 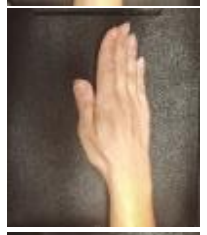 | Forearm supination (FS) |
| 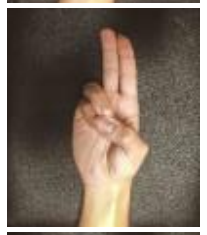 | Index and middle finger extension (IMFE) | 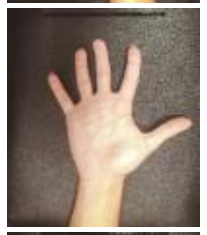 | Hand open (HO) |
| 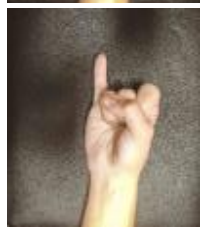 | Little finger extension (LFE) | 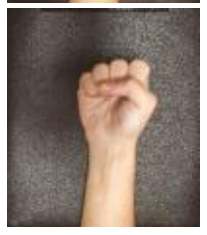 | Hand close (HC). |

## C. Experimental Protocol

After the completion of instrumentation setup, the participant is seated comfortably on the chair with both their upper limbs in a resting position. Visual instructions for performing the gestures were provided on the computer screen placed in front of the participants. The following 16 hand and wrist gestures were included in the current study (presented in Table III): Lateral prehension (LP), thumb adduction (TA), thumb and little finger opposition (TLFO), thumb and index finger opposition (TIFO), thumb and little finger extension (TLFE), thumb and index finger extension (TIFE), index and middle finger extension (IMFE), little finger extension (LFE), index finger extension (IFE), thumb extension (TE), wrist flexion (WF), wrist extension (WE), forearm supination (FS), forearm pronation (FP), hand open (HO), and hand close (HC). The order of the 16 gestures was randomized, and resting (REST) trial was collected after all 16 gestures were performed once. A ten-seconds relaxing period was provided between each trial. One continuous data acquisition of 17 gestures (including the REST) is called one run. Seven runs were performed by each subject, resulting in 119 trials or contractions (17 x 7). The subject could also request additional rest when he/she felt necessary.

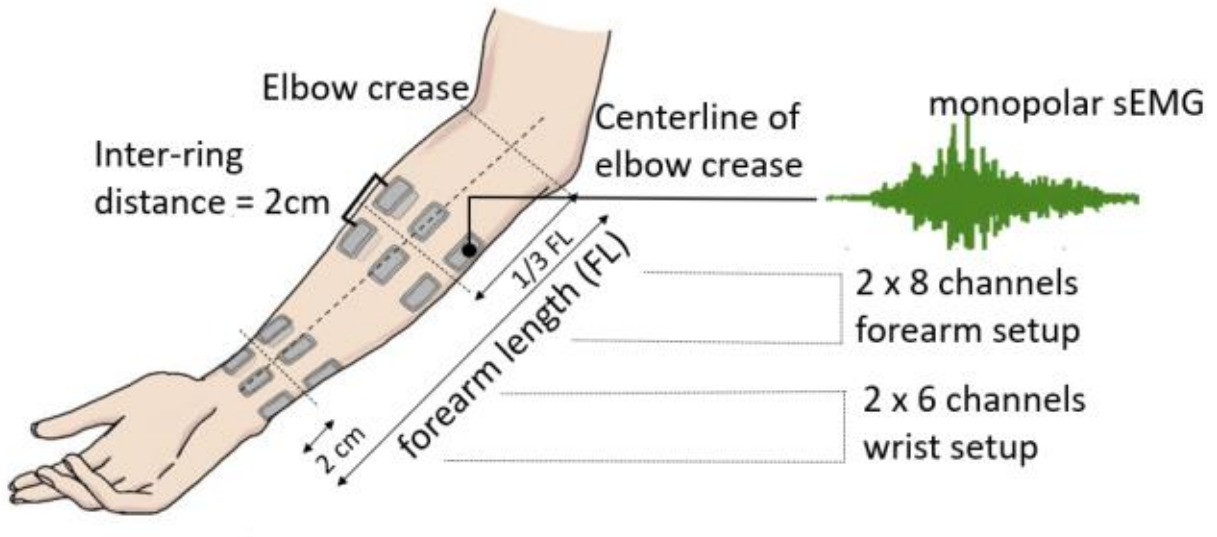

**Fig. 1 Positions of the sixteen surface electrodes on the forearm and twelve electrodes on the wrist (dorsal view). The forearm electrodes formed two rings of eight electrodes each and the wrist electrodes formed two rings of six electrodes each. The monopolar EMG of each ring was acquired for subsequent processing and analysis.**

## D. sEMG Signal Processing

For each of the proximal ring of the forearm setup (number of channels = 8) and the distal ring of the wrist setup (number of channels = 6), the monopolar sEMG signals were first re-referenced by a common average procedure. The processed signals were then segmented into 200ms width windows, with a 150 ms overlap. Each window was then processed using the frequency division technique (FDT) feature extraction [25]. This method calculates the magnitude of $L$ frequency bands. For the $i^{th}$ band, let $f_{i,1,}$ and $f_{i,E}$ denote the frequency values of the two endpoints. As such, for each window, the $i^{th}$ feature is calculated as

$$FDT_i = F\left[\sum_{j=1}^{n_i}\left|X\left(f_{i,j}\right)\right|\right], \text{ i=1,2,...}L, \quad (1)$$

where $X(\cdot)$ denotes the magnitude of the FFT spectrum, and $F[\cdot]$ denotes a non-linear transformation to obtain a smooth value for better classification results. In this study, the logarithm operator is used. In the current study, the whole EMG frequency band of EMG (20-450 Hz) is subdivided into six equal-width frequency bands: 20-92, 92-163, 163-235, 235-307, 307-378, and 378-450 Hz, consistent with prior studies [26]. Therefore, for the eight channels on the forearm and six channels on the wrist, the feature vector extracted from each window comprises of 48 and 36 FDT features from the forearm and wrist ring, respectively.

Due to the high similarity between the EMG signals from neighboring channels, the extracted features would have a high correlation. The Mahalanobis distance, which takes into consideration the correlation between features, was considered for biometric comparisons [6, 7, 11]. For a given feature vector sample $p$, which is the input from a specific user (the claimant) while performing a specific gesture, its matching score, $S_{i,j}$, with the $i$th gesture and the $j$th user, was defined as the Mahalanobis distance between the sample and the class centroid:

$$S_{i,j}(p) = \sqrt{\left(p - \mu_{i,j}\right)^{\top} \Sigma_{i,j}^{-1} \left(p - \mu_{i,j}\right)}, \quad (2)$$

where $\mu_{i,j}$ is the centroid of the gesture of the class and the user, and $\sum_{i,j}$ is the covariance matrix for the specific gesture and user class. Both the centroid and covariance were estimated from the enrollment data based on the within-day and cross-day analysis (details in Section II.G).

### *E. Multi-code Framework*

A standard biometric authentication system consists of four modules: 1) sensor module which collects the biometric data, 2) feature extractor for generating feature vectors utilized as biometric entries, 3) matcher module that compares with genuine user's template to generate a score, and 4) decision module to grant access or rejection based on a pre-set threshold. Fusion strategies at different modules for EMG-based biometric authentication have been investigated, and a fusion approached based on decision level was found to produce the best overall performance [7]. The performed hand/wrist gestures are treated as codes in the context of EMG biometrics. An example of a user's authentication code sequence of codelength $M$ can be denoted as $[C_1, C_2, ...C_m, C_M]$, where $C_m$ is the $m^{\text{th}}$ code (gesture). For the analysis below, 50 random sequences were generated by combining randomly selected six gestures out of the 16 gestures performed by each participant. A decision-level fusion of each code sequence was employed as described below [7], and the evaluation metrics were compared by varying the code length from one to six ($M$=1, 2, 3…6).

A weighted majority scheme was used for the $M$ codes of the multi-code framework. In the verification/authentication mode, the claimant user is compared to the corresponding template in the database. Hence, the authentication result is binary: 1 (for genuine user) and 0 (for impostor). For $M$ codes, let $d_{m,j}$ be the degree of certainty or the $m^{\text{th}}$ code ($C_m$) and $j^{\text{th}}$ user defined as

$$d_{m,j} = \begin{cases} 1, & C_m \text{ is correct for } j \\ 0, & otherwise \end{cases} . \quad (3)$$

The discriminant function for each user '$j$' obtained through the weighted voting is

$$g_j = \sum_{m=1}^{M} w_m \, d_{m,j}, \quad (4)$$

where $w_m$ is the weight attached to the $m^{\text{th}}$ code. This value is obtained by the single-code recognition accuracy of each gesture, averaged over the three days. Based on the value of M, $w_m$ is then normalized to 1.

$$\sum_{m=1}^{M} w_m = 1. \quad (5)$$

In the case of an authentication system, the presented user matches with the template if the discriminant $g$ has a majority (> 50%). The weight $w_m$ would be determined by the recognition accuracy of $C_m$ normalized to one.

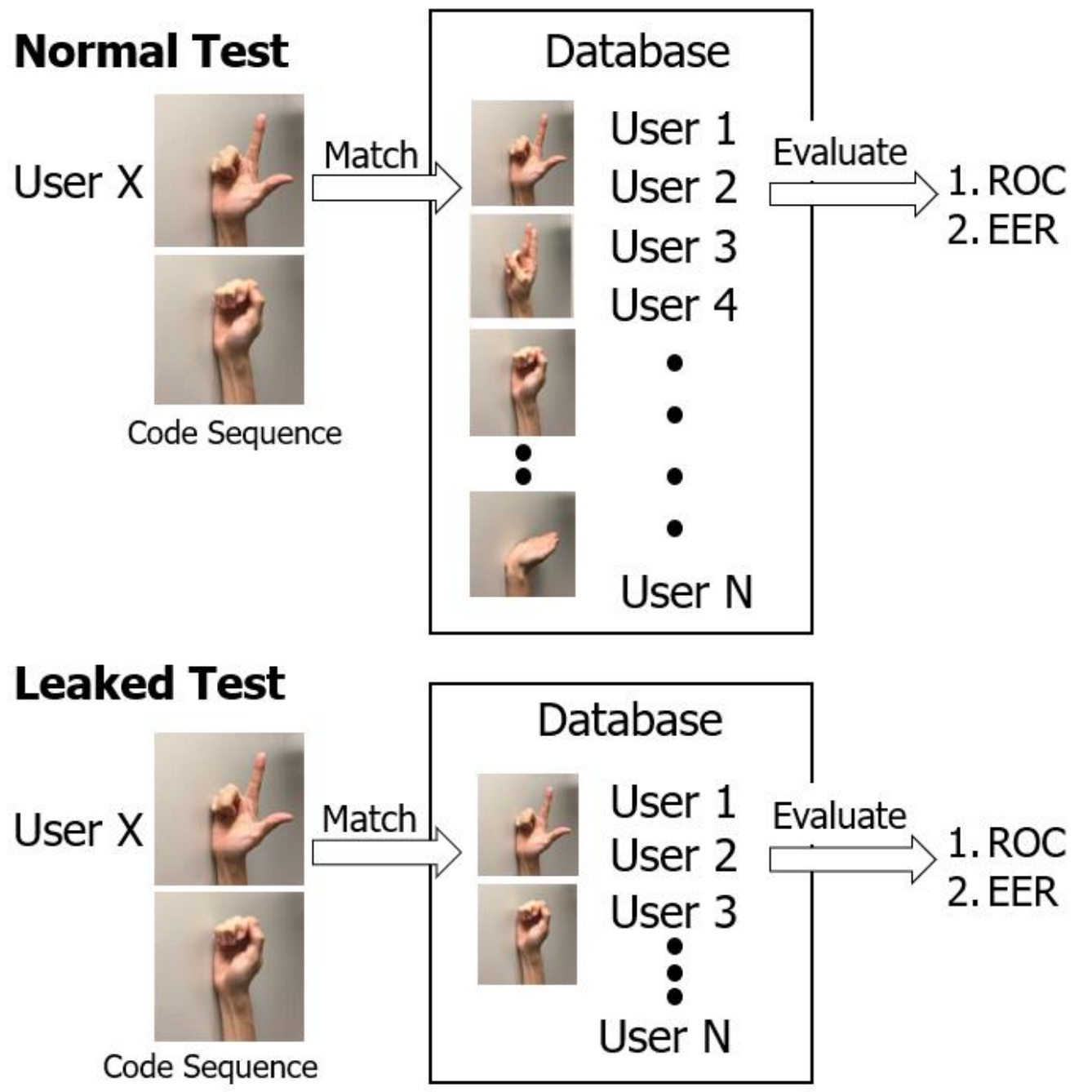

**Fig. 2 The normal test and the leaked test scenarios for biometric authentication. In both the scenarios the target user's identity is known to the claimant. The access code is unknown to the claimant in the normal test scenario, while it is known to the claimant in a leaked test scenario.**

### *F. Performance Evaluation*

The false acceptance rate (*FAR*) and false rejection rate (*FRR*) were calculated to evaluate the biometric authentication. *FAR* is the rate of accepting an impostor, and *FRR* is the rate of rejecting a genuine user. In principle, the *FAR* and *FRR* should be as small as possible for biometric applications. The detection error tradeoff curve (DEC) is the relationship between the *FAR* and *FRR*. The equal error rate (*EER*) is the point on the DEC curve where the *FAR* is equal to the *FRR*. The *EER* is a commonly used authentication metric that can be used to compare the performance of different biometric traits: The lower the *EER* value, the better the performance. For an accurate assessment of the biometric authentication capacity of the EMG biometrics, two common authentication scenarios were investigated: 1) Normal Test: where the correct code sequence was only known to the genuine user, while the impostor had no knowledge of the code sequence and presented a random sequence different from the one used by the genuine user; 2) Leaked Test: where the correct code sequence for the genuine user was compromised, and the impostor presented the correct code sequence by performing the corresponding gestures.

In both cases, for the $m$th code, the genuine score, $G_m$, was obtained from the authentication gesture $C_m$ of the genuine user. In the normal test, the impostor score $I_m$ was obtained from the other gestures performed by other users. For the leaked-test scenario, $I_m$ was obtained from $C_m$ for all the other users. Fusion schemes were implemented using $G_m$ and $I_m$ to obtain the final $FAR_M$, $FRR_M$, and $EER_M$.

## G. Within-day and cross-day analysis

In the current study, data was collected from each user over three different days comprising of seven trials each and 16 gestures in each trial. The biometric authentication performance evaluation involved a within-day analysis and two separate cross-day analyses (single cross-day and cumulative cross-day). For the within-day analysis, six trials of the gestures in each day were used as enrollment data (training) and the remaining one trial of that day was used as claimant data (testing), resulting in a leave-one-out (LOO) cross-validation scheme, equivalent to a seven-fold cross-validation. The authentication performance for each fold was estimated as described in section II.F. The cross-validation was repeated for each of the three days and the average performance metrics was reported.

For the single cross-day analysis, six trials of the gestures from one day were used as the enrollment data and the data from one trial from each of the remaining two days were used as the claimant data. This step was repeated seven times by varying the enrollment and claimant trials from the specific days and thus resulting in a between-day seven-fold cross-validation. The cross-validation was repeated three times for each day and the average authentication performance was reported.

For the cumulative cross-day analysis, six trials of the gestures from two of the three days were used as the enrollment data and the data from one trial from the remaining one day was used as the claimant data. A seven-fold cross-validation for all the seven trials was implemented by varying the enrollment and claimant trials from the specific days. The cross-validation was repeated three times for each day and the average authentication performance was reported. A graphical representation of the within-day, single cross-day and cumulative cross-day analysis is provided in Fig 3.

## H. Statistical Analysis

The study aimed to investigate the within-day and cross-day authentication performance of forearm and wrist EMG biometric system. Additionally, the performance of such a multi-code biometric systems were analyzed by varying the codelength. For each of the three analysis scenarios, *i.e.* Within-day, single cross-day and cumulative cross-day, a repeated measures ANOVA was performed on the EER of the authentication system to determine if there was any significant effect of electrode positions (two levels, *i.e.* forearm setup and wrist setup), and codelength (six levels, *i.e.* from one to six). In the case of significance in one of the factors, the levels of the other factor were sequentially fixed and the Bonferroni post-hoc analysis was performed on the significant factor. All statistical tests were performed using the RStudio 1.0. 136 (RStudio, Boston, MA).

## I. Summary of GrabMyo Dataset and sample codes

The Dataset will be made available in two open access repositories PhsioNet and IEEE dataport. For the PhysioNet database, all the EMG signal files are saved as *.dat* and *.hea* files and for the IEEE database, the signal files are saved as *.mat* files. The gesture information and the order of each trial is provided as a *.txt* file. Sample codes are provided to read the data into Matlab and Python and for implementing basic EMG signal processing. Instruction on how to use the codes are provided in the first few lines of each sample code. Additionally, a manual for data description and code usage is provided as *readme.txt* file. <link available upon request>

# III. Results

Fig. 4 shows the EER distribution in a normal test scenario while varying the codelength ($M$=1-6) for the three timeline-specific analyses: within-day, single cross-day and cumulative cross-day and two electrode setups: forearm and wrist. Fig. 5 shows the EER distribution in a leaked test scenario while varying the codelength ($M$=1-6) for the three timeline-specific analyses: within-day, single cross-day and cumulative cross-day and two electrode setups: forearm and wrist. Fig. 6 shows the ROC curves for all the different scenarios.

As expected and evident from Fig.4 and Fig. 5, within-day performance was a significantly higher ($p<0.001$) than the other two timeline-specific analyses, for both the forearm and the wrist setup. Further, the median EER of the cumulative cross-day analysis was significantly lower than the single cross-day analysis (p<0.001). The latter result was particularly encouraging, as it indicates the training data from two different days were homogeneous enough to provide significant benefits for biometric authentication, providing a basis for further incremental adaptive training. For each of the three timeline-based analyses, the effect of varying codelength and electrode setups are presented below.

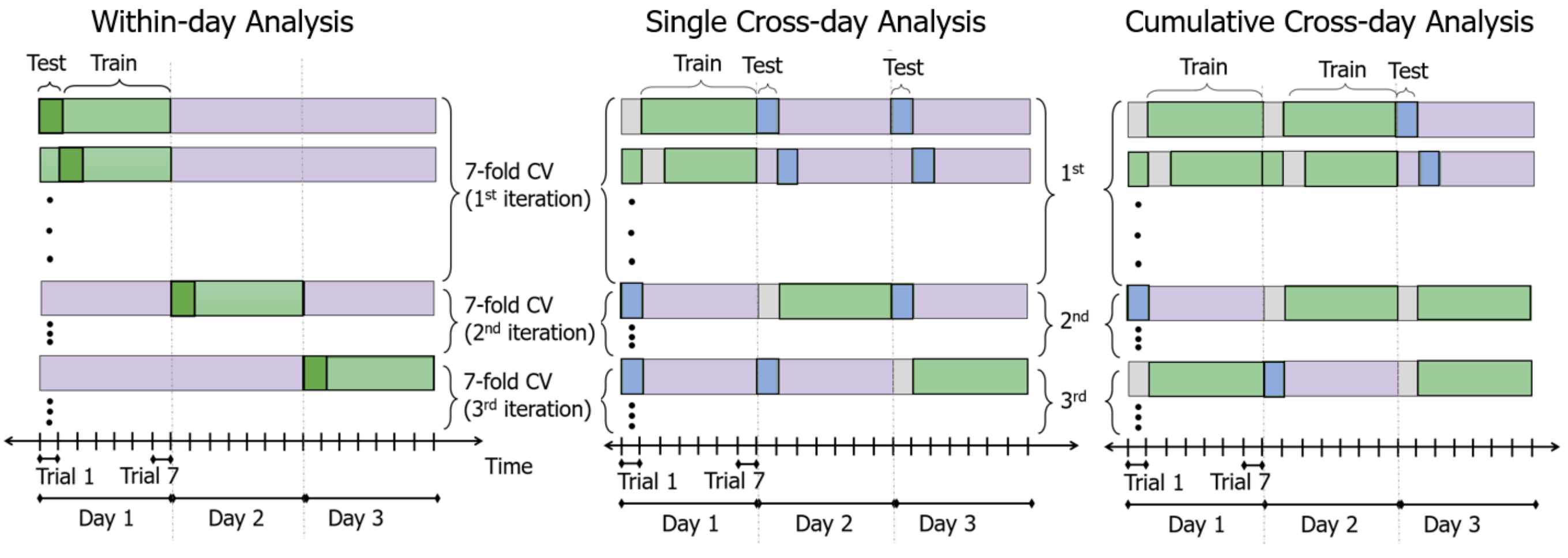

**Fig. 3 (From left to right) Within-day, Single cross-day and Cumulative cross-day analysis.** The corresponding training (enrolment) data for each analysis are represented in green; the testing (claimant) data are represented in dark green (for within-day analysis) and blue (for cross-day analysis). The x-axis represents the timeline of the study consisting of multiple days; the x-axis tick marks represent the different trials performed in one day. Each of the analysis is repeated for the three days.

### *A. Within-day analysis*

For the forearm setup, the within-day performance of the single-code (*M*=1) configuration in the normal and leaked test scenarios had a median *EER* of 0.026 (Q1=0.018, Q3=0.034) and 0.046 (Q1=0.037, Q3=0.058), respectively. These results are in agreement with a previous study recently published[7]. The performance of longer codelengths was significantly improved, as expected and the performance for *M*=3-6 reached a median value of *EER*≤0.001 for the normal test and median *EER*≤0.002 for the leaked test scenario. The single-code wrist setup had highly similar median *EER* to the corresponding forearm setup for the normal and leaked test scenarios. The best performance was obtained for codelength *M*=3-6, with a median *EER* ≤0.001 for the normal test and median *EER* ≤0.001 for the leaked test scenario.

### *B. Single cross-day analysis*

The single-code forearm setup had a median *EER* of 0.098 (Q1= 0.08, Q3=0.157) for the normal test scenario in a single cross-day analysis, which was not significantly different (p=1.0) from the median *EER* of the *M*=2 configuration (0.095, Q1=0.078, Q3=0.148). However, the median *EER* for the *M*=3 configuration (0.047, Q1=0.031, Q3=0.083)) was significantly lower (*p*<0.001) than the single-code configuration. The median *EER* was further reduced to ≤0.026 for *M*=5-6. For the leaked test scenario, the single-code forearm setup resulted in a median *EER* of 0.164 (Q1=0.134, Q3=0.239). The median *EER* significantly reduced (*p*<0.001) to 0.094 (Q1=0.062, Q3=0.161) for *M*=3 and ≤0.064 for *M*=5-6.

The wrist setup had a significantly higher (p=0.024) median *EER* than the corresponding forearm setup in the normal test scenario for all the codelengths (*M*=1-6) and the leaked test scenario for lower codelengths (*M*≤4). However, for the leaked test scenario with higher codelengths (*M*>4), there was no significant (*p*=0.16) difference between the median *EER* of the wrist setup and the forearm setup (Fig. 5). While increasing the codelength (*M*=1-6), the wrist setup had a decreasing trend of the *EER* similar to the forearm setup for both the normal and leaked test scenarios. The best performance was obtained for codelength *M*=5-6, with a median *EER* ≤0.046 for the normal test and ≤0.065 for the leaked test scenario, which was significantly lower (p<0.001) than the median *EER* for the single-code wrist setup in the normal test scenario (0.133, Q1=0.091, Q3=0.191) and the leaked test scenario (0.181, Q1=0.136, Q3=0.256), respectively.

### *C. Cumulative cross-day analysis*

The single-code forearm setup had a median *EER* of 0.069 (Q1=0.056, Q3=0.113) for the normal test scenario in a cumulative cross-day analysis. Similar to the single cross-day analysis, the median *EER* for the *M*=3 configuration (0.029, Q1=0.02, Q3=0.052) was significantly lower (*p*<0.001) than the single-code configuration. The median *EER* further reduced to ≤0.016 for *M*=5-6. For the leaked test scenario, the single-code forearm setup resulted in a median *EER* of 0.126

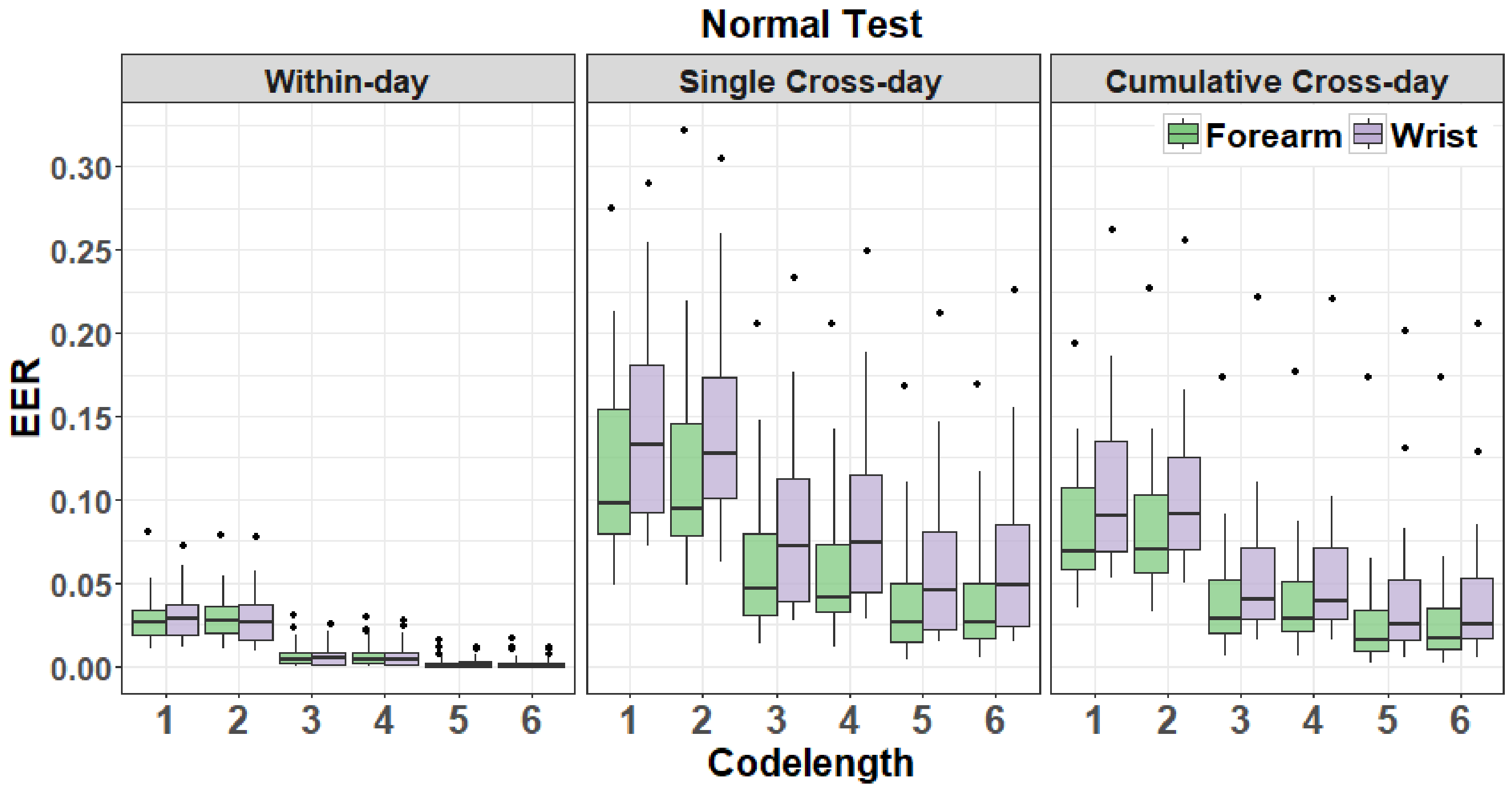

**Fig. 4 Biometric authentication performance for the Normal Test scenario.** The *EER* values for the all the individuals' forearm-setup (green) and wrist-setup (violet) in a within-day (left column), single cross-day (center column) and the cumulative cross-day (right column) analysis are shown. The x-axis represents the varying codelength (*M*=1-6). Each boxplot represents the interquartile range (IQR, $25^{th}$ – $75^{th}$ percentile) and the center horizontal line represents the median *EER* value. The whiskers (solid vertical lines) represent the datapoints within the 1.5*IQR threshold. The outliers (solid black circles) are defined as those individuals with *EER* greater than the 1.5*IQR threshold.

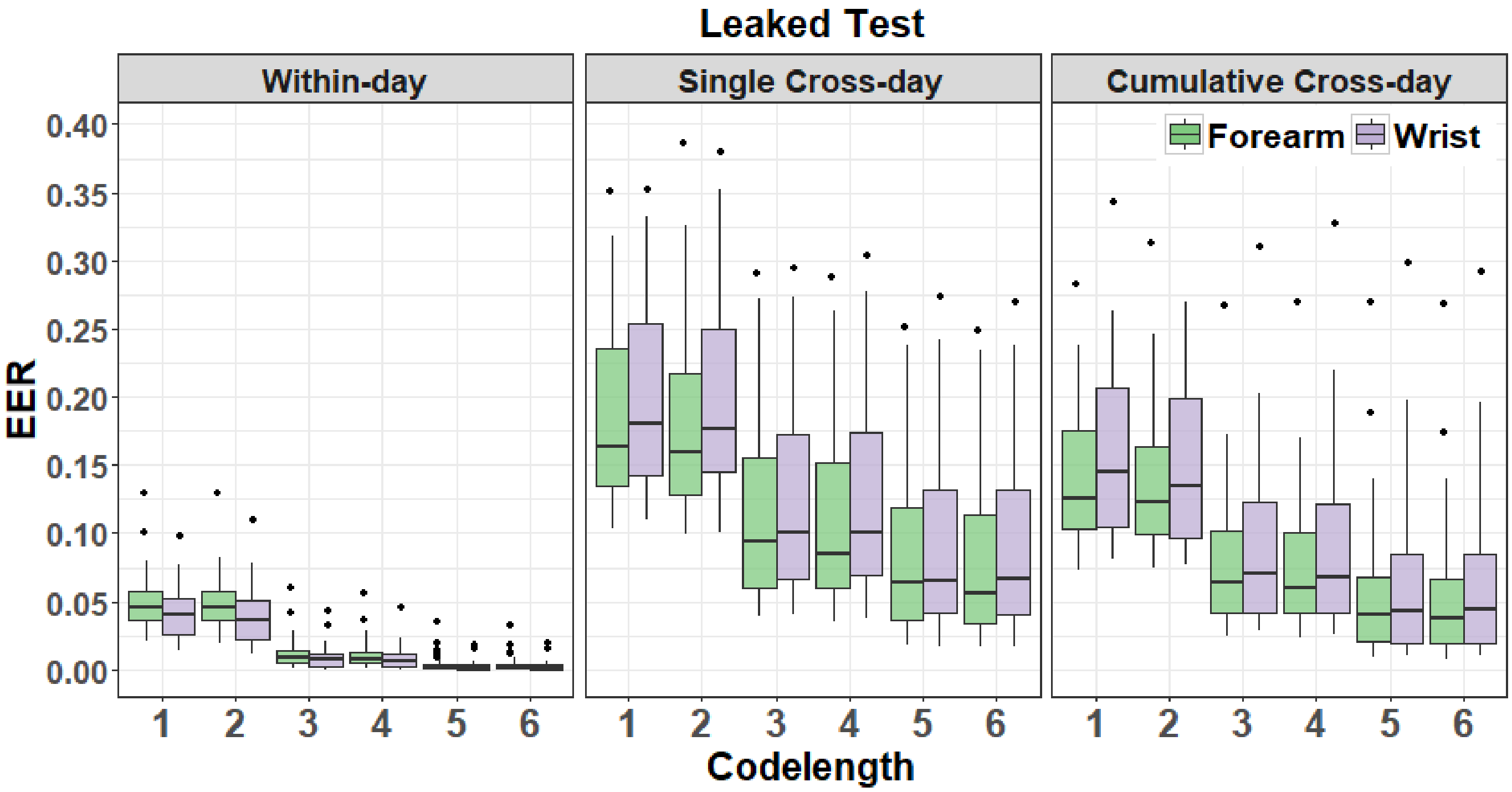

**Fig. 5 Biometric authentication performance for the Leaked Test scenario.** The *EER* values for the all the individuals' forearm-setup (green) and wrist-setup (violet) in a within-day (left column), single cross-day (center column) and the cumulative cross-day (right column) analysis are shown. The x-axis represents the varying codelength (M=1-6). Each boxplot represents the interquartile range (IQR, 25th – 75th percentile) and the center horizontal line represents the median EER value. The whiskers (solid vertical lines) represent the datapoints within the 1.5*IQR threshold. The outliers (solid black circles) are defined as those individuals whose EER was greater than the 1.5*IQR threshold.

(Q1=0.103, Q3=0.176). However, the median *EER* significantly reduced (p<0.001) to 0.064 (Q1=0.041, Q3=0.102) for *M*=3 and ≤0.041 for *M*=5-6. For both the normal and leaked test scenarios, the median *EER* values were significantly lower (*p*<0.001) than the corresponding median *EERs* for single cross-day analysis.

For all the codelengths (*M*=1-6), the wrist setup had a significantly higher (*p*<0.001) median *EER* than the corresponding forearm setup in a normal test scenario (Fig. 4). However, for the leaked test scenario and higher codelengths (*M*>2), there was no significant (*p*=0.064) difference between the median *EER* of the wrist setup and the forearm setup (Fig. 5). Similar to the forearm setup as well as the corresponding single cross-day analysis, increasing the codelength (*M*=1-6) resulted in a decreasing trend of the *EER* similar to the wrist setup for both the normal and leaked test scenarios. The best performance was obtained for codelength *M*=5-6, with a median *EER* ≤0.025 for the normal test and ≤0.043 for the leaked test scenario, which was significantly lower (*p*<0.001) than the median *EER* for the single-code wrist setup in the normal test scenario (0.09, Q1=0.069, Q3=0.136) and the leaked test scenario (0.101, Q1=0.102, Q3=0.209), respectively. Further, it was observed that for both the normal and leaked test scenarios, the median *EER* values were significantly lower (*p*<0.001) than the corresponding median *EERs* for single cross-day analysis.

## IV. DISCUSSION

The study presented the largest open-access multi-day GrabMyo dataset for biometric authentication and key benchmark metrics were demonstrated. The EMG signals were collected using electrodes from two anatomical positions: forearm and wrist, while the participants performed different hand gestures. The single-code configuration was compared to the multi-code configuration by varying the codelength from 1-6. Two evaluation scenarios, i.e., the normal and leaked test scenarios were investigated. The normal test scenario corresponds to the dual-security comprising both biometric-level (i.e. individual-specific characteristics in EMG signal) and knowledge-level (specific hand-gesture/code sequence). The leaked test scenario corresponds to the situation when the knowledge-level security (code sequence) is compromised, with only the biometric-level security is retained. A within-day and cross-day (single cross-day and cumulative cross-day) analysis was performed for each of the comparisons and the results demonstrated the biometric authentication performance, as discussed in the following sections.

TABLE IV
BIOMETRIC TRAITS CHARACTERISTICS [1]

| Biometric | Spoof-ing | Custom-izeable | Environm-ent factors | Accuracy |
|---|---|---|---|---|
| Fingerprint | Easy | No | Robust | High |
| Iris | Easy | No | Sensitive | High |
| Face | Easy | No | Sensitive | High |
| Voice | Easy | Yes | Sensitive | Low |
| Gait | Hard | Yes | Sensitive | Low |
| Keystroke Dynamics | Hard | Yes | Robust | Low |
| Biosignals | | | | |
| EEG | Hard | No | Sensitive | Low |
| ECG | Hard | No | Robust | Low |
| EMG | Hard | Yes | Sensitive | Acceptable |

## A. Single-code vs multi-code authentication performance

The single-code (or the single gesture) configuration was compared to the multi-code configuration for both the forearm and the wrist setups. It was observed that there was an overall reduction in *EER* with increasing codelengths, specifically when $M>2$. The multi-code *EER* ($M>2$) was significantly lower than the uni-code *EER* ($M>2$) for all of the comparisons. The findings were in agreement with a similar password-based study using EMG from gestures [8]. The findings in the current study were also consistent with a previous study that performed only the within-day analysis on a forearm setup [7]. For all the codelengths, it was observed that the leaked test scenario had a higher *EER* than the normal test scenario, thus suggesting that an improved performance is achieved by leveraging the knowledge-level security mode of EMG, *i.e.* with user-defined code sequences. This is a unique advantage of EMG-based biometrics as compared to other biosignals such as EEG and ECG, for which such knowledge-level security feature is not available. For the codelength $M=6$, the normal test scenario for all the analyses (within-day and cross-day) and electrode-setups (forearm and wrist) resulted in a median $EER<0.05$. For further discussions, the term 'multi-code' will only refer to the $M=6$ codelength.

## B. Within-day and Cross-day analysis

For all the codelengths, it was observed that there was a significant reduction in the median *EER* in the cross-day analysis as compared to a within-day analysis. This is expected as the EMG signals are affected by non-stationary factors such as electrode shift, skin conditions (dry vs sweat), physical conditions (rested vs exercising) [27]. While comparing the two cross-day analyses, the cumulative cross-day *EER* was significantly lower than the single cross-day *EER* for both the normal and leaked test scenarios. This suggests that EMG data from different days do have enough homogenous information such that training with data from multiple days improves the biometric authentication performance. For the cumulative cross-day normal test scenario, a multi-code configuration resulted in a median *EER* of 0.017 for the forearm setup and 0.025 for the wrist setup, which is comparable to most conventional biometrics as discussed in the below. The cross-day analysis is crucial for evaluating a biometric system as the results need to be consistent over multiple days. In the following sections, the cumulative cross-day analysis (unless specified otherwise) will be used to discuss the biometric authentication performance.

## C. Forearm vs wrist Electrode Setup

For the normal test scenario and all the codelengths ($M=1$-6), it was observed that the forearm setup had a lower EER than the wrist setup. As mentioned above, for a multi-code normal test scenario, the forearm setup had a median *EER* of 0.017, significantly lower than the median *EER* of the wrist setup 0.025. However, for the leaked test scenario and all the codelengths, it was observed that there is no significant difference between the forearm setup and the wrist setup. Specifically, for a multi-code leaked test scenario, the median *EER* of the forearm setup and wrist setup was 0.038 and 0.045, respectively. This suggests that the EMG signals from the forearm have more discriminative information than the wrist EMG signals, resulting in a better normal test biometric performance. This is also expected, as the main sources of EMG, muscle fibres, are located further away from the wrist electrodes than forearm electrodes. However, a different study found similar gesture classification performance using wrist and forearm EMG for a group of finger gestures and wrist gestures, which might be due to the difference in the nature of finger flexion/extensions [28]. For the leaked test scenario, when the gesture is the same for all the users, only the inherent EMG signal differences contribute to the biometric-level security. As acceptable biometric authentication performance ($EER<0.05$) for both the EMG setups was achieved, a wearable bracelet or wristband device could be developed for commercial applications of biometric authentication. Future research using such devices could avoid the time-consuming experimental setup and facilitate biometrics-oriented large-scale data collection.

## D. Comparison with other biometrics

Conventional biometric traits such as fingerprint and facial recognition have already been widely implemented in daily consumer applications. Extensive research has been performed on these traits, and datasets including thousands of individuals are available online [29, 30]. However, with advancing technology, data leakage and spoofing have become increasingly easier. Therefore, unconventional biometric traits such as biosignals, gait, keystroke dynamics, etc., have been investigated to address the limitations associated with the conventional biometric traits [1]. The previous studies suggest that biometric authentication performance of these traits ranges from $EER = 10^{-4}$ - 0.20 [1]. Generally, an accurate biometric authentication should achieve an EER<0.05 [find paper]. The current study found that forearm and wrist EMG-based biometrics resulted in accurate authentication using a multi-code framework for combining hand gestures. This property of customizing a code based on knowledge (behavioral traits) can be achieved using some form such as voice, gait and keystroke dynamics for improved security. This is an advantage over other biosignals such as ECG and EEG, which are highly difficult to be customizable, if possible at all [11]. Overall, the biosignals are more difficult to spoof and are one of the main indicators of liveness detection [6], a security feature requiring the user was physically present while an authentication claim is attempted. While the EMG signals are affected by multiple non-stationary factors, , the cross-day analysis suggests there are sufficient commonalities in multi-day EMG data for biometric authentication. A detailed comparison of EMG-based biometrics with the other common biometrics is provided in Table IV [1]. As EMG-based biometrics holds distinct advantages over other common biometrics, further research involving wearable devices and extensive datasets will aid its social acceptance.

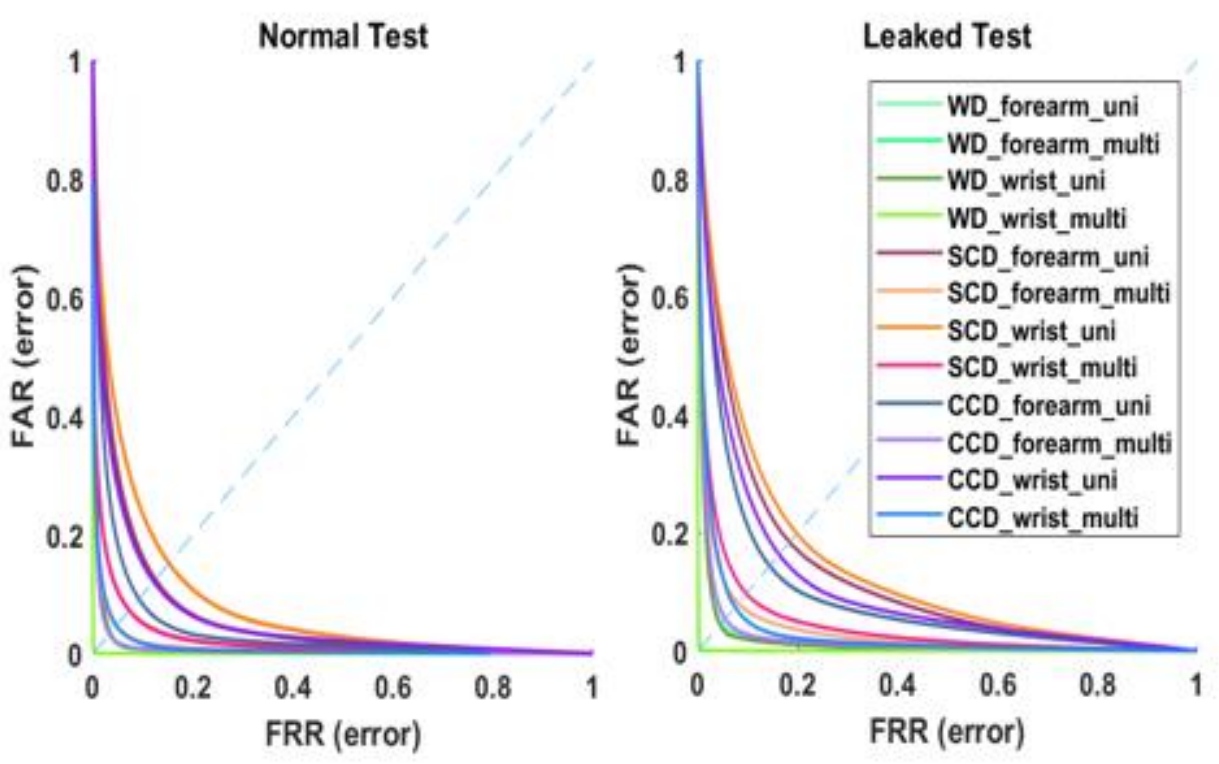

**Fig. 6 ROC curves for the two authentication scenarios.** The averaged ROC curves for the multiple analysis: within-day (WD), single cross-day (SCD) and cumulative cross-day (CCD), the two electrode setups: forearm and wrist and the two configurations: single-code (M=1) and multicode (M=6). The x axis represents the FAR values and the y axis represents the FRR values.

### E. Limitations and future research directions

While there is a clear dominance in terms of the EMG-dataset size and with high signal quality, there exists some limitation with the present study. The number of gesture codes provided to the user was limited to 16. Some single-day and multi-day open-access databases have as high as 52 gestures and 34 gestures, respectively [22, 31], although the practicality of these large numbers of gestures in the biometric context is not clear. For the present study, due to the large sample size, it was not timely feasible to collect as many gestures from all the participants. For similar reasons, multiple force levels for the same gesture could not be collected. The participant was instructed to perform the gestures at a convenient force level (as they would normally perform a gesture in their daily lives). Furthermore, no marks were left on the participants' skin for the multiple sessions. While the electrode positioning was kept consistent to the best of our ability, there might have been some electrode-shift between days. Although a limitation, such situation is in fact more realistic than tightly controlling electrode positions. It can facilitate certain electrode-shift invariant techniques as discussed below, beneficial for accurate biometric authentication. Some future research directions are:

#### 1) Improving biometric authentication

Although the multi-code biometric authentication achieved a median $EER = 0.025$ for a codelength of M = 5, practically it suggests that a user has to perform a code sequence of five gestures to achieve accurate authentication. As these gestures are short contractions, the real-life application is not entirely impossible, however, it would be beneficial to achieve higher performance with combinations of three or less gestures. Some fusion strategies, discussed in a previous multi-code EMG-based biometric research might improve the biometric authentication performance [7].

#### 2) Biometric identification

Another major biometric application is the identification mode where the system predicts the identity of the presenting user by finding the closest match. As per the definition, the identification is a more error-prone application than the authentication as the system makes $N$ comparisons, where $N$ is the number of users enrolled in the database. Therefore, the factors affecting system performance such as multiple days and sample size of the database needs to be investigated for real-life applications.

#### 3) Subject independent gesture recognition

Extensive research on EMG has been performed on gesture recognition with application in rehabilitation using prosthetic and orthotic devices, home application for assisting daily activities, virtual environment control and sign language recognition [32-34]. However, with an increase in class labels, there is exists a training burden for setting up machine learning models [35]. Recent studies have suggested deep learning techniques for cross-user calibration-free which trains generalized models using the population data, and hence reduce the training burden of the user [2, 36, 37]. The presented large-sample dataset can provide resources for such calibration-free models.

#### 4) Electrode shift-invariant techniques

One of the significant factors affecting the cross-day performance is the shift in the electrode positions. It is impossible to fix the location of armband electrodes on the forearm and wrist for daily-wear use. These variations affect the performance of both the EMG-based biometric and gesture recognition applications. Some techniques such as classification model adaptation [38, 39] and feature space transformation using transfer learning [32, 40, 41] have been suggested to address the electrode shift variations. These techniques could be further investigated to potentially improve biometrics and gesture recognition performance.

## V. Conclusion

The study presented GrabMyo, the largest multi-day EMG-based hand gestures dataset in terms of the number of recordings. The key features of the dataset included: open-access, large sample size, multi-day collection and high-quality signal acquisition. A detailed analysis was performed on the dataset for the feasibility of EMG-based biometrics as a novel trait for biometric authentications. A multi-code biometric framework had superior performance than the single-code system, suggesting enhanced security with customizable gesture passwords. Although the forearm electrode setup had a higher performance than the corresponding wrist setup, for a leaked test scenario (code-compromised), there was no difference between the two for higher codelengths ($M$>2). The worst-case scenario for both the wrist and forearm resulted in acceptable $EER$ values (<0.03), which is comparable to conventional biometrics. This could facilitate the development of wearable EMG-based bracelets and wrist bands for biometric authentication purposes. The high cross-day performance was consistent over multiple days, a necessary characteristic for biometric authentication. Therefore, the presented dataset and findings could facilitate further research on EMG-based biometrics and other gesture recognition-based applications.

## APPENDIX I

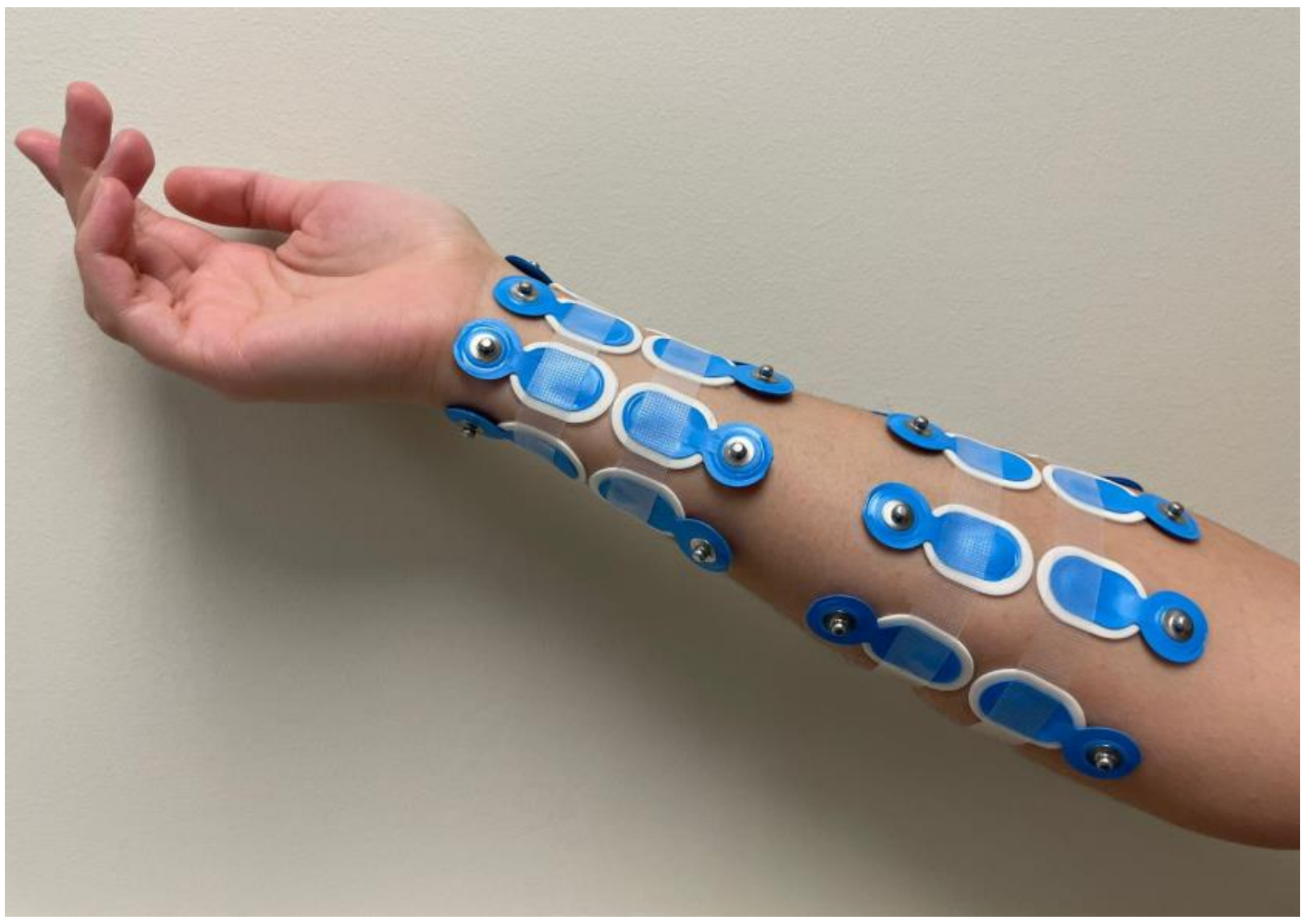

**Forearm and wrist electrode setups. The positioning of the forearm and wrist electrodes. Of the four rings, the two forearm rings consisted of 8 equispaced electrodes each and the two wrist rings included 6 electrodes each. The reference electrodes for the forearm and the wrist setups were placed on the olecranon and the styloid (hidden in this picture) respectively. To maintain consistency among participants, the first electrode on each ring was placed on the centerline connecting the anterior wrist and elbow joints and the remaining were placed in a clockwise manner.**

APPENDIX II

MULTI DAY EMG DATABASE

| Data Repository | Folders | Sub-Folders | Files | Description |
|---|---|---|---|---|
| Physionet | Session 1<br>Session 2<br>Session 3 | session***i***_subject***j*** | session***i***_subject***j***_gesture***k***_trial***l***.dat<br>session***i***_subject***j***_gesture***k***_trial***l***.hea<br>(.dat file contains 10240x32 hexadecimal values.<br>.hea file contains signal information such as sampling frequency, units and gain) | $\boldsymbol{i} \in [1,2,3]$ represents the session(day) index<br>$\boldsymbol{j} \in [1,2,\ldots.43]$ represents the subject index<br>$\boldsymbol{k} \in [1,2,\ldots.17]$ represents the gesture index<br>$\boldsymbol{l} \in [1,2,\ldots.7]$ represents the trial index |
| IEEE - Dataport | Session 1<br>Session 2<br>Session 3 | subject***i***_session***j*** | session***i***_subject***j***.mat<br>(.mat file contains 17x7 cell matrix. each cell contains 10240x32 numeric array)<br>fileinfo.txt | $\boldsymbol{i} \in [1,2,3]$ represents the session(day) index<br>$\boldsymbol{j} \in [1,2,\ldots.43]$ represents the subject index<br>*fileinfo.txt* contains the gesture definitions and their sequence. |
| Code | | | | |
| Physionet<br>IEEE - Dataport | Manual | <N/A> | readme.txt<br>fileinfo.txt<br>fileread.m (Matlab code)<br>EMGprocess.m (Matlab code)<br>fileread.ipynb (Python notebook code) | *readme.txt* general instructions on how to read datafiles<br>*fileinfo.txt* contains the gesture definitions and their sequence<br>*fileread.m* matlab code to read datafiles and convert them to numeric arrays<br>*EMGprocess.m* matlab code to perform standard EMG signal processing<br>*fileread.ipynb* python code to read datafiles and convert them to numeric arrays |

For the physionet.org repository, the signal files are converted to the waveform database (WFDB) format ( a *.dat file containing the signed 16 bit quantized value and *.hea file with the same name containing the scaling factors. For the IEEE Dataport *.mat file with same name is generated which consists of multiple recordings from trials and gestures. Data can be read into Matlab using filereader.m and into python using fileread.ipynb. A sample EMG signal processing using matlab is provided in EMGprocess.m

## Appendix III

### Normal test Analysis

| ID | WD-Uni | WD-Uni | WD-Multi | WD-Multi | SCD-Uni | SCD-Uni | SCD-Multi | SCD-Multi | CCD-Uni | CCD-Uni | CCD-Multi | CCD-Multi |
|---|---|---|---|---|---|---|---|---|---|---|---|---|
| | Fore-arm | Wrist | Fore-arm | Wrist | Fore-arm | Wrist | Fore-arm | Wrist | Fore-arm | Wrist | Fore-arm | Wrist |
| 1 | 0.030 | 0.034 | 0.003 | 0.002 | 0.085 | 0.025 | 0.123 | 0.042 | 0.062 | 0.067 | 0.018 | 0.017 |
| 2 | 0.038 | 0.049 | 0.002 | 0.003 | 0.184 | 0.064 | 0.216 | 0.123 | 0.137 | 0.189 | 0.037 | 0.067 |
| 3 | 0.023 | 0.037 | 0.001 | 0.002 | 0.077 | 0.023 | 0.093 | 0.022 | 0.049 | 0.111 | 0.010 | 0.052 |
| 4 | 0.045 | 0.060 | 0.005 | 0.007 | 0.124 | 0.043 | 0.158 | 0.070 | 0.098 | 0.125 | 0.034 | 0.053 |
| 5 | 0.027 | 0.028 | 0.006 | 0.003 | 0.130 | 0.046 | 0.230 | 0.156 | 0.101 | 0.214 | 0.035 | 0.129 |
| 6 | 0.037 | 0.048 | 0.003 | 0.004 | 0.164 | 0.075 | 0.076 | 0.015 | 0.128 | 0.149 | 0.053 | 0.065 |
| 7 | 0.034 | 0.048 | 0.001 | 0.002 | 0.150 | 0.040 | 0.159 | 0.054 | 0.095 | 0.136 | 0.017 | 0.044 |
| 8 | 0.054 | 0.061 | 0.011 | 0.010 | 0.102 | 0.031 | 0.131 | 0.046 | 0.093 | 0.122 | 0.032 | 0.052 |
| 9 | 0.035 | 0.036 | 0.002 | 0.003 | 0.166 | 0.077 | 0.199 | 0.114 | 0.131 | 0.138 | 0.066 | 0.073 |
| 10 | 0.065 | 0.072 | 0.010 | 0.012 | 0.130 | 0.044 | 0.137 | 0.032 | 0.117 | 0.146 | 0.039 | 0.069 |
| 11 | 0.020 | 0.021 | 0.000 | 0.000 | 0.097 | 0.027 | 0.134 | 0.050 | 0.064 | 0.069 | 0.012 | 0.015 |
| 12 | 0.032 | 0.022 | 0.004 | 0.001 | 0.190 | 0.119 | 0.265 | 0.153 | 0.132 | 0.135 | 0.061 | 0.055 |
| 13 | 0.024 | 0.029 | 0.001 | 0.002 | 0.061 | 0.009 | 0.127 | 0.042 | 0.041 | 0.066 | 0.004 | 0.013 |
| 14 | 0.081 | 0.067 | 0.017 | 0.012 | 0.187 | 0.086 | 0.171 | 0.082 | 0.146 | 0.262 | 0.070 | 0.205 |
| 15 | 0.042 | 0.039 | 0.002 | 0.002 | 0.158 | 0.070 | 0.138 | 0.067 | 0.113 | 0.139 | 0.037 | 0.053 |
| 16 | 0.011 | 0.013 | 0.000 | 0.000 | 0.151 | 0.056 | 0.091 | 0.020 | 0.082 | 0.070 | 0.031 | 0.017 |
| 17 | 0.014 | 0.019 | 0.000 | 0.000 | 0.215 | 0.112 | 0.114 | 0.027 | 0.143 | 0.156 | 0.066 | 0.086 |
| 18 | 0.033 | 0.029 | 0.004 | 0.002 | 0.074 | 0.011 | 0.226 | 0.156 | 0.066 | 0.091 | 0.012 | 0.025 |
| 19 | 0.032 | 0.028 | 0.001 | 0.001 | 0.085 | 0.017 | 0.220 | 0.119 | 0.067 | 0.088 | 0.011 | 0.022 |
| 20 | 0.010 | 0.007 | 0.000 | 0.000 | 0.093 | 0.026 | 0.115 | 0.059 | 0.054 | 0.069 | 0.010 | 0.017 |
| 21 | 0.032 | 0.046 | 0.001 | 0.002 | 0.102 | 0.034 | 0.139 | 0.064 | 0.068 | 0.090 | 0.019 | 0.032 |
| 22 | 0.045 | 0.038 | 0.002 | 0.001 | 0.229 | 0.119 | 0.257 | 0.137 | 0.139 | 0.162 | 0.056 | 0.070 |
| 23 | 0.019 | 0.013 | 0.000 | 0.000 | 0.050 | 0.006 | 0.087 | 0.024 | 0.036 | 0.059 | 0.002 | 0.015 |
| 24 | 0.022 | 0.023 | 0.000 | 0.000 | 0.161 | 0.052 | 0.131 | 0.032 | 0.114 | 0.139 | 0.031 | 0.051 |
| 25 | 0.014 | 0.014 | 0.000 | 0.000 | 0.057 | 0.011 | 0.081 | 0.022 | 0.041 | 0.054 | 0.004 | 0.011 |
| 26 | 0.015 | 0.015 | 0.000 | 0.000 | 0.092 | 0.026 | 0.290 | 0.226 | 0.056 | 0.061 | 0.012 | 0.014 |
| 27 | 0.030 | 0.023 | 0.001 | 0.000 | 0.080 | 0.012 | 0.191 | 0.087 | 0.061 | 0.095 | 0.006 | 0.021 |
| 28 | 0.015 | 0.018 | 0.000 | 0.000 | 0.157 | 0.040 | 0.133 | 0.034 | 0.093 | 0.116 | 0.026 | 0.040 |
| 29 | 0.023 | 0.037 | 0.001 | 0.004 | 0.091 | 0.022 | 0.140 | 0.049 | 0.068 | 0.083 | 0.012 | 0.018 |
| 30 | 0.017 | 0.021 | 0.000 | 0.000 | 0.049 | 0.004 | 0.193 | 0.091 | 0.035 | 0.053 | 0.002 | 0.005 |
| 31 | 0.024 | 0.012 | 0.001 | 0.000 | 0.275 | 0.169 | 0.072 | 0.026 | 0.194 | 0.065 | 0.173 | 0.019 |
| 32 | 0.024 | 0.032 | 0.002 | 0.003 | 0.070 | 0.020 | 0.129 | 0.063 | 0.060 | 0.086 | 0.019 | 0.036 |
| 33 | 0.042 | 0.037 | 0.002 | 0.002 | 0.098 | 0.019 | 0.138 | 0.049 | 0.076 | 0.094 | 0.011 | 0.020 |
| 34 | 0.010 | 0.007 | 0.000 | 0.000 | 0.041 | 0.005 | 0.026 | 0.001 | 0.026 | 0.036 | 0.001 | 0.003 |
| 35 | 0.026 | 0.011 | 0.000 | 0.000 | 0.089 | 0.015 | 0.073 | 0.017 | 0.069 | 0.070 | 0.007 | 0.008 |
| 36 | 0.023 | 0.016 | 0.000 | 0.000 | 0.082 | 0.007 | 0.082 | 0.022 | 0.054 | 0.087 | 0.003 | 0.012 |
| 37 | 0.017 | 0.024 | 0.000 | 0.001 | 0.107 | 0.030 | 0.171 | 0.055 | 0.069 | 0.105 | 0.017 | 0.040 |
| 38 | 0.018 | 0.021 | 0.001 | 0.001 | 0.092 | 0.025 | 0.095 | 0.027 | 0.064 | 0.082 | 0.014 | 0.025 |
| 39 | 0.032 | 0.034 | 0.003 | 0.002 | 0.080 | 0.023 | 0.087 | 0.016 | 0.062 | 0.074 | 0.014 | 0.020 |
| 40 | 0.020 | 0.023 | 0.000 | 0.000 | 0.121 | 0.048 | 0.079 | 0.017 | 0.071 | 0.087 | 0.019 | 0.026 |
| 41 | 0.028 | 0.043 | 0.001 | 0.002 | 0.127 | 0.024 | 0.129 | 0.025 | 0.092 | 0.126 | 0.014 | 0.037 |
| 42 | 0.032 | 0.029 | 0.001 | 0.001 | 0.079 | 0.016 | 0.071 | 0.011 | 0.056 | 0.080 | 0.009 | 0.022 |
| 43 | 0.010 | 0.015 | 0.000 | 0.000 | 0.032 | 0.002 | 0.218 | 0.119 | 0.022 | 0.031 | 0.001 | 0.001 |
| **Q1** | **0.018** | **0.018** | **0.000** | **0.000** | **0.080** | **0.091** | **0.016** | **0.024** | **0.056** | **0.069** | **0.010** | **0.017** |
| **M** | **0.026** | **0.028** | **0.001** | **0.001** | **0.098** | **0.133** | **0.026** | **0.049** | **0.069** | **0.090** | **0.017** | **0.025** |
| **Q3** | **0.034** | **0.038** | **0.002** | **0.003** | **0.157** | **0.191** | **0.051** | **0.085** | **0.209** | **0.136** | **0.035** | **0.053** |

WD: Within Day, SCD: Single Cross Day, CCD: Cumulative Cross Day

Uni- Single-code (Codelength = 1), Multi-code (Codelength = 6) Q1= $25^{th}$ percentile, M= Median, Q3 =$75^{th}$ percentile

## Appendix IV

### Leaked test Analysis

| ID | WD-Uni | WD-Uni | WD-Multi | WD-Multi | SCD-Uni | SCD-Uni | SCD-Multi | SCD-Multi | CCD-Uni | CCD-Uni | CCD-Multi | CCD-Multi |
|---|---|---|---|---|---|---|---|---|---|---|---|---|
| | Fore-arm | Wrist | Fore-arm | Wrist | Fore-arm | Wrist | Fore-arm | Wrist | Fore-arm | Wrist | Fore-arm | Wrist |
| 1 | 0.045 | 0.047 | 0.002 | 0.003 | 0.130 | 0.168 | 0.038 | 0.063 | 0.095 | 0.130 | 0.024 | 0.044 |
| 2 | 0.061 | 0.068 | 0.003 | 0.004 | 0.256 | 0.286 | 0.117 | 0.202 | 0.207 | 0.266 | 0.072 | 0.188 |
| 3 | 0.037 | 0.046 | 0.002 | 0.002 | 0.126 | 0.116 | 0.033 | 0.017 | 0.082 | 0.094 | 0.012 | 0.012 |
| 4 | 0.068 | 0.077 | 0.007 | 0.007 | 0.192 | 0.215 | 0.071 | 0.107 | 0.149 | 0.199 | 0.050 | 0.091 |
| 5 | 0.049 | 0.048 | 0.012 | 0.005 | 0.201 | 0.332 | 0.093 | 0.270 | 0.162 | 0.343 | 0.057 | 0.292 |
| 6 | 0.054 | 0.066 | 0.002 | 0.003 | 0.228 | 0.111 | 0.111 | 0.015 | 0.189 | 0.097 | 0.075 | 0.010 |
| 7 | 0.061 | 0.071 | 0.004 | 0.004 | 0.240 | 0.239 | 0.101 | 0.115 | 0.167 | 0.210 | 0.048 | 0.081 |
| 8 | 0.081 | 0.072 | 0.013 | 0.004 | 0.156 | 0.163 | 0.044 | 0.045 | 0.141 | 0.149 | 0.040 | 0.037 |
| 9 | 0.060 | 0.055 | 0.004 | 0.005 | 0.269 | 0.273 | 0.163 | 0.187 | 0.224 | 0.234 | 0.133 | 0.131 |
| 10 | 0.101 | 0.091 | 0.018 | 0.015 | 0.200 | 0.175 | 0.094 | 0.040 | 0.178 | 0.163 | 0.068 | 0.038 |
| 11 | 0.045 | 0.028 | 0.001 | 0.000 | 0.177 | 0.197 | 0.071 | 0.081 | 0.126 | 0.145 | 0.038 | 0.051 |
| 12 | 0.055 | 0.029 | 0.008 | 0.002 | 0.298 | 0.352 | 0.212 | 0.238 | 0.223 | 0.215 | 0.121 | 0.145 |
| 13 | 0.038 | 0.038 | 0.002 | 0.003 | 0.103 | 0.162 | 0.018 | 0.047 | 0.072 | 0.128 | 0.009 | 0.032 |
| 14 | 0.130 | 0.098 | 0.033 | 0.019 | 0.285 | 0.256 | 0.175 | 0.148 | 0.235 | 0.209 | 0.141 | 0.105 |
| 15 | 0.061 | 0.053 | 0.002 | 0.002 | 0.243 | 0.195 | 0.122 | 0.101 | 0.175 | 0.162 | 0.059 | 0.064 |
| 16 | 0.018 | 0.024 | 0.000 | 0.000 | 0.239 | 0.136 | 0.134 | 0.032 | 0.146 | 0.088 | 0.094 | 0.013 |
| 17 | 0.024 | 0.028 | 0.000 | 0.000 | 0.320 | 0.180 | 0.237 | 0.066 | 0.252 | 0.140 | 0.174 | 0.046 |
| 18 | 0.053 | 0.040 | 0.005 | 0.004 | 0.116 | 0.293 | 0.021 | 0.236 | 0.103 | 0.210 | 0.017 | 0.125 |
| 19 | 0.058 | 0.040 | 0.002 | 0.001 | 0.159 | 0.320 | 0.054 | 0.235 | 0.120 | 0.279 | 0.031 | 0.197 |
| 20 | 0.020 | 0.009 | 0.000 | 0.000 | 0.156 | 0.162 | 0.054 | 0.084 | 0.101 | 0.086 | 0.023 | 0.016 |
| 21 | 0.047 | 0.059 | 0.001 | 0.001 | 0.151 | 0.184 | 0.042 | 0.078 | 0.103 | 0.152 | 0.019 | 0.056 |
| 22 | 0.064 | 0.048 | 0.003 | 0.001 | 0.334 | 0.346 | 0.238 | 0.219 | 0.238 | 0.233 | 0.125 | 0.089 |
| 23 | 0.046 | 0.026 | 0.001 | 0.000 | 0.113 | 0.151 | 0.028 | 0.053 | 0.087 | 0.109 | 0.016 | 0.032 |
| 24 | 0.040 | 0.027 | 0.000 | 0.000 | 0.231 | 0.183 | 0.101 | 0.047 | 0.179 | 0.125 | 0.066 | 0.023 |
| 25 | 0.032 | 0.021 | 0.000 | 0.000 | 0.111 | 0.125 | 0.026 | 0.031 | 0.085 | 0.101 | 0.011 | 0.020 |
| 26 | 0.031 | 0.015 | 0.001 | 0.000 | 0.164 | 0.332 | 0.056 | 0.263 | 0.108 | 0.218 | 0.026 | 0.207 |
| 27 | 0.053 | 0.028 | 0.001 | 0.000 | 0.135 | 0.251 | 0.029 | 0.113 | 0.106 | 0.209 | 0.015 | 0.077 |
| 28 | 0.024 | 0.014 | 0.000 | 0.000 | 0.253 | 0.169 | 0.117 | 0.056 | 0.174 | 0.128 | 0.092 | 0.023 |
| 29 | 0.036 | 0.048 | 0.002 | 0.003 | 0.134 | 0.181 | 0.032 | 0.060 | 0.113 | 0.131 | 0.019 | 0.046 |
| 30 | 0.038 | 0.031 | 0.000 | 0.000 | 0.108 | 0.271 | 0.017 | 0.158 | 0.084 | 0.188 | 0.008 | 0.078 |
| 31 | 0.044 | 0.019 | 0.002 | 0.000 | 0.351 | 0.110 | 0.248 | 0.045 | 0.283 | 0.073 | 0.268 | 0.017 |
| 32 | 0.041 | 0.051 | 0.003 | 0.005 | 0.118 | 0.189 | 0.040 | 0.108 | 0.101 | 0.154 | 0.032 | 0.075 |
| 33 | 0.068 | 0.047 | 0.004 | 0.002 | 0.165 | 0.207 | 0.047 | 0.097 | 0.137 | 0.172 | 0.030 | 0.068 |
| 34 | 0.023 | 0.010 | 0.000 | 0.000 | 0.088 | 0.046 | 0.010 | 0.003 | 0.060 | 0.034 | 0.004 | 0.001 |
| 35 | 0.042 | 0.014 | 0.001 | 0.000 | 0.146 | 0.105 | 0.041 | 0.026 | 0.123 | 0.080 | 0.027 | 0.014 |
| 36 | 0.047 | 0.021 | 0.001 | 0.000 | 0.144 | 0.118 | 0.035 | 0.034 | 0.109 | 0.087 | 0.017 | 0.012 |
| 37 | 0.037 | 0.036 | 0.000 | 0.001 | 0.191 | 0.230 | 0.075 | 0.074 | 0.144 | 0.159 | 0.046 | 0.036 |
| 38 | 0.036 | 0.036 | 0.001 | 0.003 | 0.161 | 0.148 | 0.068 | 0.046 | 0.121 | 0.121 | 0.041 | 0.030 |
| 39 | 0.057 | 0.042 | 0.006 | 0.002 | 0.139 | 0.127 | 0.043 | 0.029 | 0.114 | 0.113 | 0.028 | 0.020 |
| 40 | 0.038 | 0.035 | 0.001 | 0.001 | 0.216 | 0.120 | 0.124 | 0.028 | 0.145 | 0.094 | 0.063 | 0.016 |
| 41 | 0.050 | 0.052 | 0.002 | 0.001 | 0.216 | 0.177 | 0.070 | 0.041 | 0.176 | 0.136 | 0.050 | 0.023 |
| 42 | 0.057 | 0.050 | 0.003 | 0.001 | 0.137 | 0.115 | 0.030 | 0.017 | 0.106 | 0.087 | 0.019 | 0.008 |
| 43 | 0.013 | 0.020 | 0.000 | 0.000 | 0.054 | 0.307 | 0.004 | 0.224 | 0.039 | 0.204 | 0.001 | 0.124 |
| **Q1** | **0.037** | **0.026** | **0.001** | **0.000** | **0.134** | **0.136** | **0.033** | **0.040** | **0.103** | **0.101** | **0.019** | **0.002** |
| **M** | **0.046** | **0.040** | **0.002** | **0.001** | **0.164** | **0.181** | **0.056** | **0.066** | **0.126** | **0.145** | **0.038** | **0.044** |
| **Q3** | **0.058** | **0.052** | **0.004** | **0.004** | **0.239** | **0.256** | **0.119** | **0.144** | **0.113** | **0.136** | **0.068** | **0.089** |

WD: Within Day, SCD: Single Cross Day, CCD: Cumulative Cross Day

Uni- Single-code (Codelength = 1), Multi-code (Codelength = 6) Q1= 25$^{th}$ percentile, M= Median, Q3 =75$^{th}$ percentile